\documentclass[conference]{IEEEtran}

\usepackage[utf8]{inputenc}
\usepackage{amsmath,amsthm,amssymb}
\usepackage{cite}
\usepackage{hyperref}
\usepackage{graphicx} 

\newtheorem{theorem}{Theorem}

\newtheorem{definition}{Definition}

\usepackage{amsthm}
\usepackage{enumitem}
\newtheorem{remark}{Remark}
\usepackage{algorithm}
\usepackage[noend]{algpseudocode}
\usepackage[]{balance}
\usepackage{booktabs} 

\begin{document}
	
\title{On the Impossibility of a Perfect Hypervisor}	
	\author{
		\IEEEauthorblockN{Mordechai Guri}
		\IEEEauthorblockA{Ben-Gurion University of the Negev, Israel \\
			Email: gurim@post.bgu.ac.il}
	}
	
	\maketitle

\begin{abstract}
	We establish a fundamental impossibility result for a \emph{perfect hypervisor}—one that
	(1) preserves every observable behavior of any program exactly as on bare metal and
	(2) adds zero timing or resource overhead.
	
	Within this model we prove two theorems.
	
	\noindent\textbf{(1) Indetectability Theorem.}
	If such a hypervisor existed, \emph{no guest-level program, measurement, or timing test}
	could distinguish it from native execution; all traces, outputs, and timings would be
	identical.
	
	\noindent\textbf{(2) Impossibility Theorem.}
	Despite that theoretical indetectability, a perfect hypervisor cannot exist on any
	machine with finite computational resources.
	
	These results are architecture-agnostic and extend beyond hypervisors to any
	virtualization layer—emulators, sandboxes, containers, or runtime-instrumentation
	frameworks.  Together they provide a formal foundation for future work on the
	principles and limits of virtualization.
\end{abstract}

\begin{IEEEkeywords}
	virtualization, hypervisor, virtual machine monitor, impossibility result,
	timing overhead, detection, nesting, resource constraints
\end{IEEEkeywords}
	
	\section{Introduction}
	Virtualization underpins much of modern computing, allowing one physical 
	machine to host multiple \emph{guest} operating systems or processes via 
	a \emph{hypervisor}, also known as a \emph{virtual machine monitor} (VMM). 
	Engineers have sought to minimize the overhead and visibility of 
	virtualization using hardware extensions (Intel VT-x \cite{intel-vtx}, AMD-V  \cite{amd-v}), 
	paravirtualization, and other optimizations. These efforts have led to 
	considerable improvements but have not closed the gap to the 
	hypothetical \emph{perfect hypervisor}: 
	\emph{one that neither changes observable behavior nor exact timing} 
	compared to running on bare metal.
	
	In this paper, we prove two fundamental results regarding this ideal:
	
	\begin{itemize}
		\item \textbf{Indetectability of a Perfect Hypervisor:} 
		If such a perfect hypervisor existed, it would be impossible for 
		any program to detect its presence, since all execution traces, 
		I/O results, and timing would be \emph{identical} to those on 
		bare metal.
		\item \textbf{Impossibility of a Perfect Hypervisor:} 
		No such hypervisor can actually exist on a finite-resource machine, 
		as shown by a contradiction argument involving infinite nesting.
	\end{itemize}

	\subsection{Contributions}
	\begin{enumerate}
		\item \emph{Perfect Hypervisor Definition.} Based on the foundational virtualization framework by \emph{Popek and Goldberg} \cite{popek1974formal}, we define a hypothetical \emph{perfect hypervisor} that strictly satisfies two conditions: (1) it preserves every observable behavior of a program exactly as if it were executing on bare metal, and (2) it introduces zero additional timing or resource overhead.	
		\item We present and prove an \emph{Indetectability Theorem} showing that, under these strong conditions, the hypothetical perfect hypervisor would be fundamentally undetectable by any algorithmic test.
		\item We present and prove an \emph{Impossibility Theorem} demonstrating that such a perfect hypervisor cannot exist in any physically realizable system with finite resources.
	\end{enumerate}
	
	We provide proofs grounded in fundamental principles of computation rather than architectural or technological limitations, highlighting that the impossibility arises inherently at the conceptual level.
	
	\subsection{Paper Organization}
The paper is organized as follows. Section~\ref{sec:model} defines the system model, timing semantics, and the notion of a perfect hypervisor. Section~\ref{sec:indetectability} proves that such a hypervisor would be fundamentally undetectable. Section~\ref{sec:impossibility-finite} establishes its impossibility under finite-resource constraints. Section~\ref{sec:discussion} discusses implications for nested and hardware-assisted virtualization. Section~\ref{sec:conclusion} concludes.

	\section{System Model and Definitions}
	\label{sec:model}
	
	\subsection{Machine Model, Execution, and I/O}
	We assume a physical machine $\mathcal{M}$ with a finite set of 
	resources: CPU cycles, memory (including cache hierarchy), and I/O devices. 
	Time is discretized into \emph{ticks}, each representing an indivisible 
	quantum (e.g., a clock cycle or minimal scheduler slice). A 
	\emph{bare-metal} execution of a program $P$ on $\mathcal{M}$ can thus 
	be represented by a sequence of state transitions
	\[
	s_0 \;\xrightarrow{1}\; s_1 \;\xrightarrow{1}\; 
	s_2 \;\xrightarrow{1}\;\cdots\;\xrightarrow{1}\; s_{t},
	\]
	where each step consumes exactly one tick and yields a new system 
	state $s_i$. These states include the contents of memory, registers, 
	and any relevant I/O buffers.

	More formally, a machine state~$s_i$ is defined as a tuple:
	\[
	s_i = (\texttt{Mem}_i, \texttt{Reg}_i, \texttt{IO}_i),
	\]
	where:
	\begin{itemize}
		\item $\texttt{Mem}_i: \mathbb{A} \rightarrow \mathbb{V}$ is the memory state, modeled as a mapping from memory addresses ($\mathbb{A}$) to values ($\mathbb{V}$).
		\item $\texttt{Reg}_i: \mathbb{R} \rightarrow \mathbb{V}$ is the register state, mapping register identifiers ($\mathbb{R}$), including general-purpose registers, special-purpose registers, the program counter, and flags, to values ($\mathbb{V}$).
		\item $\texttt{IO}_i$ captures the state of I/O buffers and devices, whenever applicable, including pending operations and device statuses.
	\end{itemize}

	Each state transition updates these components deterministically based on the executed instruction or event occurring within that tick.

	Each instruction or operation in $P$ may also produce \emph{observable 
		results}, such as output to a display, network packets, or data written 
	to storage. For convenience, we denote all such \emph{I/O events} 
	collectively as a set of time-stamped outputs 
	$\{\,(e_1, t_1), (e_2, t_2), \dots\}$, where $t_i$ is the 
	tick at which event $e_i$ occurred.
	
	\subsection{Hypervisor (Virtual Machine Monitor)}
	\label{sec:def-hypervisor}
	Following the virtualization model of Popek and Goldberg~\cite{popek1974formal},
	a \emph{virtual machine monitor} (VMM)—commonly referred to today as a \emph{hypervisor}—acts
	as an intermediary layer that presents a virtual machine $\mathcal{V}$ to the guest program~$P$.
	
	Formally, $H$ transforms the bare-metal state transitions 
	into transitions that \emph{appear} identical to a direct execution 
	on physical hardware, but behind the scenes, $H$ may intercept 
	privileged instructions, manage memory mappings, handle device I/O, 
	and schedule CPU access.
	
	If $H$ itself runs on $\mathcal{M}$, the combined sequence of states 
	from the perspective of $P$ can be abstractly written as:
	\[
	\mathcal{M}: \quad 
	s_0 \;\xrightarrow{H}\; s_1 \;\xrightarrow{H}\; s_2 
	\;\xrightarrow{H}\;\cdots\;\xrightarrow{H}\; s_{t}.
	\]
	Here, each transition may consume some number of \emph{machine} ticks, 
	depending on how $H$ emulates or traps the guest's requests.
	
	\subsection{Timing and Overhead: Partial Native Execution}
	\label{sec:timing-overhead}
	Modern virtualization often executes \emph{some} instructions natively 
	without trap-and-emulate overhead (e.g., via Intel VT-x, AMD-V). 
	Hence, \emph{not all} instructions necessarily incur hypervisor overhead. 
	Nonetheless, a \emph{perfect hypervisor} must ensure that 
	\emph{no step}---including privileged instructions, memory or device I/O, 
	interrupt handling, etc.---takes longer than it would on bare metal. 
	
	\begin{definition}[Timing Overhead]
		\label{def:overhead}
		Let $T_{\mathcal{M}}(P,n)$ be the number of \emph{ticks} used by program $P$ 
		to complete $n$ operations on bare metal $\mathcal{M}$. 
		Let $T_{H}(P,n)$ be the number of ticks used by $P$ under 
		hypervisor $H$ to complete the same $n$ operations. 
		We define the timing overhead as
		\[
		\text{Overhead}(H,P,n) \;=\; T_{H}(P,n) \;-\; T_{\mathcal{M}}(P,n).
		\]
		A hypervisor has \emph{zero overhead} if and only if
		\[
		\text{Overhead}(H,P,n) = 0
		\quad \forall P \in \mathcal{P},\; \forall n \in \mathbb{N}.
		\footnotemark
		\]
		\footnotetext{That is, for all possible (i.e., computable) programs expressible on a Turing-complete machine.}
		where $\mathcal{P}$ denotes the space of all computable programs. 
		Equivalently, 
		\[
		\forall P \in \mathcal{P},\; \forall n \in \mathbb{N},\quad T_{H}(P,n) = T_{\mathcal{M}}(P,n).
		\]
		Even if some instructions are not executed natively, \emph{every} instruction and event 
		must preserve exact timing equivalence.
	\end{definition}

	One can also view overhead at the level of individual instructions:
	\[
	\text{Overhead}(H,P,i) 
	\;=\; \sum_{j=1}^i 
	\Bigl(\,t_{H}(P,j) \;-\; t_{\mathcal{M}}(P,j)\Bigr),
	\]
	where \(t_{H}(P,j)\) and \(t_{\mathcal{M}}(P,j)\) are the ticks to run the 
	\(j\)-th instruction under $H$ vs.\ bare metal, respectively.  
	A perfect hypervisor requires this sum to be zero for \emph{every} prefix 
	\(1 \le i \le n\).
	
	\subsection{Behavioral Equivalence}
	\begin{definition}[Behavioral Equivalence]
		\label{def:behavioral-equivalence}
		A hypervisor $H$ provides \emph{behaviorally equivalent} execution 
		if, for every program $P$ and every operation index $i$ 
		($1 \le i \le n$), the \emph{observable state} and the 
		\emph{I/O events} produced by $P$ at step $i$ under $H$ 
		match exactly those produced when $P$ runs on bare metal for $i$ steps. 
		Formally, if $s_i^{\mathcal{M}}$ denotes the system state on bare metal 
		after $i$ steps and $s_i^H$ denotes the hypervisor-based state, then 
		\(\forall i,\ s_i^{H} = s_i^{\mathcal{M}}\) in all observables, 
		and the I/O event history is identical up to step $i$.
	\end{definition}
	
	\subsection{Perfect Hypervisor}
	\begin{definition}[Perfect Hypervisor]
		\label{def:perfect}
		A hypervisor $H$ on $\mathcal{M}$ is called \emph{perfect} if:
		\begin{enumerate}
			\item \textbf{Behavioral Equivalence:} $H$ is behaviorally equivalent 
			to bare metal for all programs $P$ (per 
			Definition~\ref{def:behavioral-equivalence}).
			\item \textbf{Zero Timing Overhead:} 
			$\text{Overhead}(H,P,n)=0$ for all programs $P$ and all $n$ 
			(Definition~\ref{def:overhead}).
		\end{enumerate}
	\end{definition}
	
	In simpler terms, a perfect hypervisor never changes either 
	the time it takes to run $P$ up to any step or the external 
	outputs produced by $P$ at each step.

	\subsection{Observationally Identical Execution.}
	Based on Definitions~\ref{def:overhead} and \ref{def:behavioral-equivalence}, we say that the execution of a program $P$ under a hypervisor $H$ 
	is \emph{observationally identical} to its execution on bare metal $\mathcal{M}$ if, 
	for every step $i$, the externally visible system state $s_i^H$ equals $s_i^{\mathcal{M}}$, 
	and the time $T_H(P,i)$ to reach that state matches $T_{\mathcal{M}}(P,i)$. 
	This means that $P$ observes the same outputs, timings, and side effects under $H$ 
	as it would on physical hardware. Consequently, if a hypervisor is \emph{perfect} 
	(Definition~\ref{def:perfect}), it guarantees observationally identical execution 
	for all programs and all execution prefixes.

	\section{Indetectability of a Perfect Hypervisor}
	\label{sec:indetectability}

\begin{theorem}[Indetectability Theorem]
	\label{thm:indetectability2}
	If $H$ is a perfect hypervisor on $\mathcal{M}$, then for \emph{any} detection
	algorithm $D$ \emph{executing inside the virtual machine context exposed by~$H$},
	the execution of $D$ under $H$ is observationally identical to the execution of
	$D$ on bare metal.  Consequently, \emph{no} such detector can reliably
	distinguish $H$ from $\mathcal{M}$.
\end{theorem}

\subsection{Proof}
Let $D$ be an arbitrary detection algorithm—or “detector”—\emph{executing entirely
	inside the virtual-machine context provided by $H$}.
Its goal is to decide whether it is running on bare metal or under a hypervisor.  
By assumption of a perfect hypervisor, we have two primary invariants:
\emph{behavioral equivalence} and \emph{zero timing overhead}.  Formally,
	
	\[
	\forall i: \quad s_i^H = s_i^{\mathcal{M}} 
	\quad (\text{behavioral equivalence}),
	\]
	\[
	\forall i: \quad T_H(D,i) = T_{\mathcal{M}}(D,i)
	\quad (\text{timing equivalence}),
	\]
	
	where
	\begin{itemize}[leftmargin=*]
		\item $s_i^{H}$ is the externally visible state of $D$
		(including any guest-accessible side-channel measurements)
		at step~$i$ when running under~$H$;
		\item $s_i^{\mathcal M}$ is the externally visible state of $D$
		at the same step when running directly on the bare-metal
		machine~$\mathcal M$; and
		\item $T_H(D,i)$ and $T_{\mathcal M}(D,i)$ denote the cumulative
		ticks (or other resource metrics) observable at step~$i$
		in the respective runs.
	\end{itemize}
	
	Because these two runs are observationally equivalent at every step $i$, the entire 
	trace of externally visible states, outputs, and timings accessible to $D$ is 
	identical in both environments. Formally, let 
	
	\[
	\tau_H \;=\; \bigl\{ (s_1^H,\,T_H(D,1)),\;\dots,\;(s_n^H,\,T_H(D,n)) \bigr\}
	\]
	and
	\[
	\tau_{\mathcal{M}} \;=\; \bigl\{ (s_1^{\mathcal{M}},\,T_{\mathcal{M}}(D,1)),\;\dots,\;(s_n^{\mathcal{M}},\,T_{\mathcal{M}}(D,n)) \bigr\}
	\]
	be the full execution traces (states and timings) of $D$ under $H$ and on 
	bare metal, respectively. From the above invariants, we have 
	$\tau_H = \tau_{\mathcal{M}}$. 
	
	Hence, for any function
	\[
	f \;:\; \Bigl(\text{States}\times\text{Timings}\Bigr)^n \;\longrightarrow\; \mathcal{R},
	\]
	that $D$ might compute from its observable execution trace (including any final 
	decision bit “hypervisor vs.\ bare metal”), it follows that
	\[
	f\bigl(\tau_H\bigr) \;=\; f\bigl(\tau_{\mathcal{M}}\bigr).
	\]
	In other words, $f$ \emph{must return the same value in both runs} because $f$ is 
	evaluated on identical input data in each environment.
	
	Therefore, $D$’s outcome does not differ between bare metal and hypervisor execution, 
	implying it cannot exceed random guessing in determining whether $H$ is present. 
	Thus, no detection algorithm $D$ can reliably distinguish $H$ from $\mathcal{M}$.
	
\begin{center}
	\fbox{\parbox{0.95\linewidth}{\textbf{A perfect hypervisor is fundamentally undetectable by any guest-level algorithm.}}}
\end{center}
\hfill $\qed$

\begin{remark}
	This result implies that \emph{any} approach to detect the presence of a “perfect” 
	hypervisor must fail with probability at least as high as random guessing. Such a hypervisor would be 
	\emph{intrinsically undetectable}, as it offers no measurable or inferable artifact 
	to the guest environment.
\end{remark}

\section{Impossibility of a Perfect Hypervisor on a Finite-Resource Machine}
\label{sec:impossibility-finite}

\begin{theorem}[No Perfect Hypervisor on Finite Resources]
	\label{thm:no-perfect2}
	\emph{Claim:} Let $\mathcal{M}$ be a physical machine with \emph{finite} resources
	(e.g., memory, CPU cycles in a bounded interval, or finite I/O bandwidth).
	There exists no hypervisor $H$ on $\mathcal{M}$ that is 
	\emph{perfect} for \emph{all} programs $P$, under the definitions established 
	in Definitions~\ref{def:behavioral-equivalence}--\ref{def:perfect}.
\end{theorem}

\subsection{Proof}

\subsubsection{Step 1: Assume Existence of a Perfect Hypervisor}
\label{sec:assumption-perfect-H}
\noindent
\textbf{Assumption.}
Suppose, for contradiction, that there is a hypervisor $H$ on $\mathcal{M}$ 
satisfying both conditions in Definition~\ref{def:perfect} for \emph{any} program $P$.  
We will show that such $H$ cannot coexist with the finiteness of $\mathcal{M}$’s resources.

\subsubsection{Step 2: Arbitrary Self-Nesting}
\label{sec:self-nesting}
\noindent
\textbf{Rationale.}
Under condition (1), $H$ is \emph{behaviorally} indistinguishable from $\mathcal{M}$.  
Hence, if a guest program attempts to install another copy of $H$ (i.e., run the hypervisor code within itself), $H$ must permit it exactly as if running on real hardware.  
Refusal to do so would be observable to the guest, violating behavioral equivalence.

\medskip
\noindent
\textbf{Construction.}
Define $H^1 = H$, and for each integer $k \ge 1$, let $H^{k+1}$ be an instance of the 
\emph{same} hypervisor code $H$ running as a guest under $H^k$.  
By condition (2) of \eqref{def:perfect} (Zero Timing Overhead), 
each $H^{k+1}$ also perceives no additional latency or scheduling delay.  
Formally, for every $k$ and every step $i$:
\[
s_i^{H^k} \;=\; s_i^{\mathcal{M}} 
\quad\text{and}\quad
T_{H^k}(P,i) \;=\; T_{\mathcal{M}}(P,i).
\]
Thus, we can \emph{nest} $H$ arbitrarily many times without introducing any 
detectable difference compared to bare metal.

\subsubsection{Step 3: Contradiction from Finite Resources}
\label{sec:finite-resources-contradiction}

\begin{figure}[]
\centering
\includegraphics[width=0.6\linewidth]{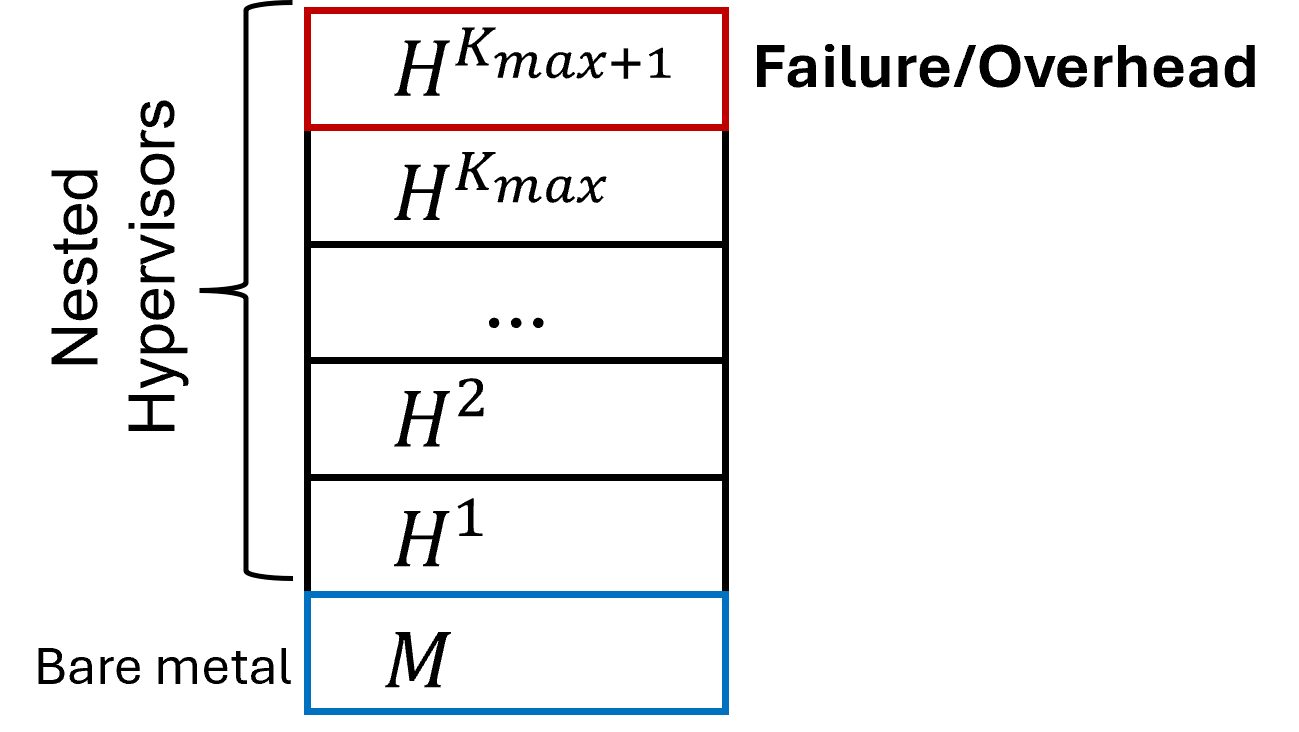}

\caption{Conceptual illustration of nested virtualization under the perfect-hypervisor assumption. Layers $H^{1}$ through $H^{k_{\max}}$ must each appear cost-free to the guest. Instantiating the $(k_{\max}+1)$-st layer exhausts finite resources, causing either an observable failure or non-zero overhead; either outcome violates \textit{Definition}~\ref{def:perfect}.}
\label{fig:nesting-contradiction}
\end{figure}

\noindent
Because $\mathcal{M}$ is explicitly stated to have finite memory and CPU capacity, 
the possibility of creating infinitely many nested hypervisors $H^1, H^2, \dots$ 
leads to a direct conflict (Figure \ref{fig:nesting-contradiction}).

We make this precise below:

\begin{enumerate}[label=\alph*)]
	\item \textbf{Positive Resource Consumption.}  
	Every hypervisor instance, however small or optimized, consumes a strictly positive 
	amount of memory and other hardware resources (e.g., control structures, scheduling 
	overhead for intercepts, device virtualization data).  
	Let $\mu_k > 0$ represent the memory footprint of $H^k$.  
	The total memory consumed by $k$ nested instances is:
	\[
	M(k)\;=\;\sum_{j=1}^k \mu_j \;\ge\; k \cdot \mu_{0},
	\]
	where $\mu_0>0$ is a lower bound on $H$’s minimal memory usage. Since $\mathcal{M}$ 
	has finite memory $M_{\max}$, there must be some integer $k_{\max}$ beyond which the 
	machine cannot instantiate any additional hypervisor layer (i.e., 
	$M(k_{\max}+1) > M_{\max}$).
	
	\item \textbf{Observable Failure or Resource Multiplexing Overhead.}  
	Once we reach nesting depth $k_{\max}+1$, the system faces two outcomes, 
	each contradicting one of the perfectness properties in \eqref{def:perfect}:
	
	\begin{itemize}
		\item \textbf{Case 1: Failure or Refusal to Launch.}  
		If the new hypervisor instance $H^{k_{\max}+1}$ cannot be created due to 
		memory exhaustion, then the guest sees an error or denial 
		absent on true hardware (which would manifest a 
		hardware fault in a specific manner that $H$ cannot perfectly mask).  
		This violates the \emph{behavioral equivalence} requirement (1).
		
		\item \textbf{Case 2: Attempted “Compression,” “Reallocation,”, “Simulation,” etc.}  
		Suppose $H$ tries to mask the resource exhaustion by strategies such as compressing memory, reallocating CPU time, or by simulating the \emph{appearance} of a successful $H^{k_{\max}+1}$ launch without actually instantiating it. In all cases, $H$ performs additional internal operations that do not occur on bare metal in response to the same guest instructions.
		
		\begin{itemize}
			\item \emph{Timing overhead is introduced.}  
			Compression or emulation of nested behavior, as well any other algorithm, must consume CPU cycles or I/O bandwidth. 
			If $P$ experiences even small delays (e.g., longer TLB misses, slower I/O), then condition (2) in \eqref{def:perfect} is violated:  
			\[
			T_H(P,i) \ne T_{\mathcal{M}}(P,i).
			\]
			
			\item \emph{Visible state changes occur.}  
			To mask the presence of $H^{k_{\max}+1}$, $H$ must intercept control transfers, emulate system calls, or simulate internal VM state. These introduce inconsistencies — e.g., incorrect exception codes, unexpected cache behavior, or changes in page table layout — that $P$ can detect. This violates behavioral equivalence:  
			\[
			s_i^H \ne s_i^{\mathcal{M}}.
			\]
		\end{itemize}
		
		In either sub-case, the hypervisor deviates from the bare-metal reference model, and can no longer remain perfect under 
		Definitions~\ref{def:behavioral-equivalence}--\ref{def:perfect}.
	\end{itemize}
\end{enumerate}

\vspace{0.5em}
\noindent\textbf{Conclusion of the Contradiction.}
We see that continuing to nest $H$ forever on a finite-resource $\mathcal{M}$ 
eventually produces an \emph{observable} distinction in either states or timings, 
contrary to \eqref{def:perfect}.  
Hence, our initial assumption that $H$ can be \emph{perfect} for all programs $P$ must be false.  
\emph{No hypervisor} can satisfy both behavioral equivalence and zero overhead 
on a finite-resource machine. 

\medskip
\noindent
\begin{center}
\fbox{\parbox{0.95\linewidth}{\textbf{No perfect hypervisor exists on a finite-resource machine.}}}
\end{center}
 \hfill $\qed$
 
 \section{Discussion}
 \label{sec:discussion}
 
\subsection{Generalization Beyond Hypervisors}
 While our proofs explicitly address the impossibility of a \emph{perfect hypervisor}, the underlying reasoning naturally extends to broader classes of virtualization technologies. Common abstractions beyond hypervisors—such as emulators, sandboxes, containers, runtime instrumentation frameworks, and other execution-isolation environments—share fundamental properties that make the generalization straightforward.
 
 More formally, we observe that any virtualization or isolation layer that satisfies analogous conditions to those described in Definitions~\ref{def:behavioral-equivalence} and~\ref{def:perfect}—particularly behavioral equivalence and zero timing overhead—must confront the same resource-constraint limitations and contradictions discussed in Section~\ref{sec:impossibility-finite}. Specifically, the following reasoning applies broadly:
 
 \begin{enumerate}[label=(\roman*)]
 	\item \textbf{Emulators and Full-System Simulators:} Emulators translate and simulate instructions from a guest architecture to a host architecture. Achieving perfect behavioral equivalence and timing fidelity would necessitate precise simulation of CPU pipelines, caches, timing interrupts, and other hardware features at zero overhead. As shown previously, finite resources inherently impose limits on such precise, overhead-free emulation, making a \emph{perfect emulator} similarly unattainable.
 	
 	\item \textbf{Sandboxes and Containers:} These lighter-weight virtualization mechanisms typically isolate processes or applications by controlling their system calls, network interfaces, or file accesses. Even these seemingly lightweight solutions inevitably require nonzero overhead (e.g., syscall interception, memory isolation mechanisms, cgroup or namespace management), and attempting to completely hide this overhead violates the finite-resource constraint argument. Thus, a sandbox or container environment achieving both perfect equivalence and zero overhead would face precisely the contradictions described earlier.
 	
 	\item \textbf{Runtime Instrumentation and Debugging Environments:} Instrumentation frameworks (e.g., Valgrind, DynamoRIO, PIN) inject additional code or trap instructions to monitor or modify program behavior. Any instrumentation step necessarily adds overhead. Therefore, maintaining observational indistinguishability with uninstrumented execution, particularly in timing-critical applications, is fundamentally impossible under the conditions discussed.
 	
 \end{enumerate}
 
 In general, our impossibility result reflects an intrinsic limitation rooted not only in hypervisor technology, but in the concept of virtualization itself. Any software-based abstraction layer placed between hardware and executing programs must inevitably either manifest measurable resource-consumption overhead or fail to reproduce behavioral equivalence with bare-metal execution faithfully. Consequently, the impossibility theorem established in this paper extends naturally and rigorously to all virtualization and isolation layers seeking absolute indistinguishability and zero overhead.

 \medskip
 \noindent
 \begin{center}
 	\fbox{\parbox{0.95\linewidth}{\textbf{Perfect virtualization is fundamentally impossible under finite-resource constraints.}}}
 \end{center}

 \subsection{From Technical to Fundamental Analysis}
 
 While many technical studies have extensively analyzed the practical aspects of virtualization overhead, detection techniques, and architecture-specific vulnerabilities~\cite{Garfinkel2007, Ferrie2006, Rutkowska2007}, this paper intentionally abstracts away from implementation-specific artifacts. Instead, we provide foundational theorems establishing the intrinsic theoretical limits of virtualization itself, independent of the underlying architecture, implementation details, or specific virtualization technology. This approach ensures generalizability, clarity, and rigor, laying the groundwork for deeper theoretical and practical insights into virtualization systems.

 	\subsection{Nested Virtualization}
 Real systems that allow nested virtualization (e.g., KVM on KVM, 
 or VMware on VMware) accumulate overhead with each layer. 
 Our result clarifies why: 
 \emph{some} overhead is inherent to all layers; 
 otherwise, infinite nesting would be possible with no cost, 
 violating resource constraints.
 
 \subsection{Hardware-Assisted Virtualization}
 While CPU extensions (Intel VT-x, AMD-V) reduce overhead for many 
 virtualization operations (often letting non-privileged instructions 
 run at native speed), they cannot eliminate overhead 
 \emph{entirely}. Theorem~\ref{thm:no-perfect2} applies equally to 
 any hardware-assisted approaches, since even these must handle 
 certain privileged instructions, resource mappings, and interrupts.

\section{Conclusion}
\label{sec:conclusion}
We have shown that a \emph{perfect hypervisor}—one that incurs \emph{no timing or
	resource overhead} and exhibits \emph{no observable behavioral differences}—would
be \emph{undetectable by any guest‐level algorithm}
(Theorem~\ref{thm:indetectability2}).  We then proved that such a hypervisor
cannot exist on a finite-resource machine (Theorem~\ref{thm:no-perfect2}); allowing
infinite, cost-free nesting would contradict basic resource constraints.

This boundary underscores a broader lesson: overhead can be minimized but never
completely eliminated across all programs and execution prefixes.  Each
additional layer of nesting must accumulate some cost, so any claim of “perfect
virtualization” must reckon with the impossibility formalised here.

	\balance
	\bibliographystyle{IEEEtran}

\end{document}